\pdfoutput=1
\documentclass[fleqn,usenatbib]{mnras}

\usepackage{mathptmx}

\usepackage[T1]{fontenc}
\usepackage{ae,aecompl}

\pdfoutput=1 
\usepackage{savesym}
\usepackage{amsmath}
\savesymbol{iint}
\usepackage{txfonts}
\restoresymbol{TXF}{iint}
\usepackage{natbib, graphicx}
\usepackage{color}
\usepackage{hyperref}
\usepackage{booktabs}
\usepackage{pdflscape}
\usepackage{multirow}
\usepackage{longtable}
\usepackage{pdflscape}
\usepackage{threeparttable}
\usepackage{subcaption}
\usepackage{float}
\usepackage{amssymb}
\usepackage{adjustbox}

\title{On the iron ionization balance of cool stars}

\author[M. Tsantaki]{
M. Tsantaki,$^{1}$\thanks{E-mail: \href{mailto:Maria.Tsantaki@astro.up.pt}{Maria.Tsantaki@astro.up.pt}}
N. C. Santos,$^{1}$
S. G. Sousa,$^{1}$ 
E. Delgado-Mena,$^{1}$
V. Adibekyan,$^{1}$ \\
\newauthor{D. T. Andreasen$^{1}$}
\\
$^{1}$Instituto de Astrof\'isica e Ci\^encias do Espa\c{c}o, Universidade do Porto, CAUP, Rua das Estrelas, Porto, 4150-762, Portugal}
\date{Accepted XXX. Received YYY; in original form ZZZ}

\pubyear{2018}
     
\begin{document}
\label{firstpage}
\pagerange{\pageref{firstpage}--\pageref{lastpage}}
\maketitle

\begin{abstract}
High-resolution spectroscopic studies of solar-type stars have revealed higher iron abundances derived from singly ionized species compared to neutral, violating the ionization equilibrium 
under the assumption of local thermodynamic equilibrium. In this work, we investigate the overabundances of \ion{Fe}{ii} lines reported in our previous work for a sample of 451 solar-type HARPS 
stars in the solar neighborhood. The spectroscopic surface gravities of this sample which emerge from the ionization balance, appear underestimated for the K-type stars. In order to understand this 
behavior, we search our \ion{Fe}{ii} line list for unresolved blends and outliers. First, we use the VALD to identify possible unresolved blends around our lines and calculate which ones are strong 
enough to cause overestimations in the equivalent width measurements. Second, for our sample we use reference parameters (effective temperature and metallicity) and the \textit{Gaia} DR2 
parallaxes to derive surface gravities (trigonometric gravities) and calculate the \ion{Fe}{i} and \ion{Fe}{ii} abundances from different line lists. We exclude the \ion{Fe}{ii} lines which produce 
overabundances above 0.10\,dex. The derived surface gravities from the clean line list are now in agreement with the trigonometric. Moreover, the difference between \ion{Fe}{i}and \ion{Fe}{ii} 
abundance does not show now a correlation with the effective temperature. Finally, we show that the ionization balance of Ti can provide better estimates of surface gravities than iron. With 
this analysis, we provide a solution to the ionization balance problem observed in the atmospheres of cool dwarfs. 

\end{abstract}



\section{Introduction}\label{intro}

A standard method to determine the atmospheric parameters via spectroscopy is to measure the equivalent widths (EW) of several spectral lines of a metallic species and calculate their 
abundances. Then, we derive the effective temperature ($T_{\mathrm{eff}}$) and surface gravity ($\log g$) of a star when excitation and ionization balances are satisfied simultaneously 
under the assumption of local thermodynamic equilibrium (LTE). In most cases, we use neutral and singly ionized iron lines because they are numerous and well studied in terms of their 
atomic data in the optical wavelength region. This method has been successfully applied to a plethora of studies for FGK-type stars from the characterization of Galactic stellar 
populations \citep[e.g.][]{Adibekyan2012} to the characterization of planet-host stars \citep[e.g.][]{santos13} mainly with high resolution spectroscopy. Even though the above methods 
provide high enough precision and accuracy for $T_{\mathrm{eff}}$ and metallicity\footnote{We note that in this study we use iron as a proxy for the overall metallicity which 
is defined as: $[Fe/H]$ $\equiv$ $\log$ $\frac{N_{Fe}}{N_{H}}_{\star}$ -- $\log$ $\frac{N_{Fe}}{N_{H}}_{\odot}$, where N is the number of atoms per unit volume. 
Iron abundance is defined as: $\log A(Fe)$ $\equiv$ $\log$ $\frac{N_{Fe}}{N_{H}}_{\star}$ + 12.}, the determination of surface gravity shows some caveats \citep[see][for surface gravity 
comparisons between different methods]{morel2012, Mortier2014} and is difficult to constrain from the ionization 
balance of iron lines because the neutral-to-singly-ionized ratio is not very sensitive to gravity changes compared to other ionization 
levels \citep[see][]{Gray2005}. Moreover, the number of ionized lines is very limited compared to neutral lines in the optical. 

Several works in high resolution report discrepancies between neutral and singly ionized iron abundances for dwarf stars in several open clusters (Hyades, Pleiades, and M34) and the 
Ursa Major moving group without considering non-LTE effects \citep{yong2004, schuler2006, chen2008, schuler2010, aleo2017}. The discrepancies are stronger for the cooler stars 
($T_{\mathrm{eff}} \lesssim$\,5200\,K) with systematic higher \ion{Fe}{ii} abundance over \ion{Fe}{i}. The enhanced \ion{Fe}{ii} abundances over \ion{Fe}{i} can be quite significant, 
for example in the high resolution study of Pleiades the difference reaches up to 0.8\,dex \citep{schuler2010}. The deviation from the ionization balance for the cooler dwarfs is also 
present in field stars in the solar neighborhood \citep{allende2004, ramirez07, bensby2014}. In this work, we refer as the \textit{ionization balance problem}, the observed 
differences between \ion{Fe}{i} and \ion{Fe}{ii} abundances in LTE and not to the LTE departures when the mean intensity, J$_{\nu}$, is larger than the Planck function, 
B$_{\nu}$, for species in their minority ionization stage in the ultraviolet wavelengths \citep[e.g.][]{asplund2005}.

Departures from the LTE were suggested as responsible for the \ion{Fe}{i}--\ion{Fe}{ii} discrepancy affecting the correct $\log g$ determination for very metal-poor stars 
\citep{korn2003,ruchti2012} but they are not expected to occur around solar metallicity K-type dwarfs \citep{lind2012}.  In addition, other model uncertainties related to granulation and 
activity of K-type stars have been proposed to explain these differences in the case of HIP\,86400 \citep{ramirez2008} but to be fully understood 3D non-LTE investigations should be carried out 
\citep{amarsi2016}.

Alternatively, if the adopted stellar parameters are not correct, the iron abundances will not be correct either. For example, if we increase the 
$T_{\mathrm{eff}}$ of a  K-type star by 100\,K which is a typical error, this leads to a reduction of the \ion{Fe}{ii} abundance by $\sim$0.15\,dex but leaves the \ion{Fe}{i} abundance 
almost unchanged. 

The ionization imbalance affects directly the determination of surface gravity when it is derived from the \ion{Fe}{i}--\ion{Fe}{ii} tuning. The comparison of surface gravity from 
spectroscopy and estimated from more direct methods such as using astrometry (trigonometric $\log g$), shows spurious correlations with $T_{\mathrm{eff}}$ with higher deviations 
for the cooler stars \citep{tsantaki13, Tabernero2017, delgado2017}. In the \textit{Gaia} era, we have precise parallaxes for a huge number of stars, and therefore, 
trigonometric gravities can be used as reference, assuming well constrained $T_{\mathrm{eff}}$, to test the accuracy of spectroscopic $\log g$.

This spectroscopic method relies significantly on the quality of the iron line list and has to be carefully selected. In previous work, we showed that blended iron lines can 
produce overestimations in the spectroscopic effective temperature scale compared to the photometric one \citep[][hereafter TS13]{tsantaki13}. Also, \cite{aleo2017} recently showed 
that a careful selection of their \ion{Fe}{ii} lines can decrease the \ion{Fe}{i}--\ion{Fe}{ii} differences for K-type stars in the Hyades cluster but not eliminate them. 

In this work, we use high resolution spectra of a well-studied sample of solar-type dwarfs to investigate if the observed overabundances of \ion{Fe}{ii} in the 
literature arise from the quality of the iron line list. If this is the case for the \ion{Fe}{i}--\ion{Fe}{ii} discrepancy, other mechanisms are not 
necessary, at least in first order, to explain the ionization balance problem. Under this assumption, a selection of optimal iron lines to solve the ionization 
balance problem will also eliminate the differences between the spectroscopic and trigonometric gravities and their trends with $T_{\mathrm{eff}}$. 

\section{The iron line list}\label{linelist}

The compilation of the iron line list is very important for the abundance analysis, especially for the cooler stars because their spectra are highly line crowded and the EW of these spectral lines 
cannot be measured accurately. Our line list is taken from TS13 which was visually checked to discard blended lines in the cooler stars. The effective temperature and metallicity derived with 
the TS13 line list have been compared with other literature sources and have been validated for their accuracy. For example, the comparison of the $T_{\mathrm{eff}}$ from TS13 with more 
model independent methods, such as the infrared flux method (IRFM) and interferometry showed very good agreement, and the metallicity is in agreement with \cite{sousa2008}. We note that 
the metallicity in TS13 is derived only from \ion{Fe}{i} lines and the $T_{\mathrm{eff}}$ from the excitation balance of the \ion{Fe}{i} lines. Therefore, the \ion{Fe}{i} lines of this 
list are reliable enough to estimate effective temperatures and metallicities. However, the TS13 line list failed to provide accurate surface gravities in particular for the cooler stars 
($T_{\mathrm{eff}}$<\,5200\,K). As mentioned before, we obtain surface gravity with this method by forcing \ion{Fe}{i} and \ion{Fe}{ii} abundances to be equal. We suggest that the 
\ion{Fe}{ii} lines must mostly contribute to the wrong surface gravity estimates we observe. Therefore, we search for unresolved blends around the \ion{Fe}{ii} lines in the TS13 line list 
which contains 120 \ion{Fe}{i} and 18 \ion{Fe}{ii} lines. The \ion{Fe}{i} lines are more numerous than \ion{Fe}{ii} and the latter are more difficult to be found isolated. Also, because of 
their small number their dispersion in abundances is high and usually a 3\,$\sigma$ clipping of outliers which is a commonly used does not work.

In the following Sections, we perform two separate tests to select the optimal \ion{Fe}{ii} lines for our abundance analysis.

\subsection{Searching for blends in VALD}\label{vald}

With this test we want to discover unresolved blends which cannot be detected neither by visual inspection nor by our automated tools. In this case, these lines will be 
fitted by a single Gaussian instead of two (or more), leading to an overestimation of their true EW. In our works, we use \texttt{ARES}\footnote{\texttt{ARES\,2.0}: 
\href{http://www.astro.up.pt/~sousasag/ares}{http://www.astro.up.pt/~sousasag/ares}} \citep{sousa2015} for the automatic EW measurements which has been extensively used in the literature for high resolution studies 
\citep[e.g. for the \textit{Gaia}-ESO survey,][]{Smiljanic2014, jofre2014}. The suggested distance of two consecutive lines to be resolved as separate with \texttt{ARES} is more than 
0.07\,\AA{} for high resolution spectra and specifically, this value has been set for the analysis of the HARPS GTO planet search sample \citep{sousa2008, sousa2011}. Ideally, this value 
should be set depending on spectral type, instrumental resolution, and signal-to-noise ratio but since it is difficult to evaluate this value per spectrum, we use the default value provided for 
\texttt{ARES} in this work. This test is based on the idea that since \texttt{ARES} cannot resolve two lines in shorter distance than 0.07\,\AA{}, the unresolved blends will be inside 
the intervals of 0.14\,\AA{} wide. To be conservative, we select wider intervals, 0.20\,\AA{} wide, and we query around our \ion{Fe}{ii} lines within these intervals for all possible 
lines from the Vienna Atomic Line Database\footnote{VALD: \href{http://vald.inasan.ru/~vald3/php/vald.php}{http://vald.inasan.ru/~vald3/php/vald.php}} \citep[VALD;][]{piskunov95, kupka99, Ryabchikova2015}. 

VALD includes dozens of lines in our regions to be potential blends. However, not all lines that appear inside the intervals lead to overestimation of the iron EW because their strengths 
depend on the physical conditions in the atmospheres they are formed. For example, ionic metallic lines are stronger in the atmospheres of the hotter stars. 

We calculate the theoretical EW for all the VALD lines to find which ones are strong enough relative to the EW of the \ion{Fe}{ii} line in each interval. For this analysis, we use the 
atomic data (oscillator strengths, $\log gf$) from VALD, the spectral analysis code, \texttt{MOOG}\footnote{\texttt{MOOG\,2017}: \href{http://verdi.as.utexas.edu/moog.html}{http://verdi.as.utexas.edu/moog.html}}
\citep[\texttt{ewfind} driver,][]{sneden1973, Sneden2012}, and a \cite{kurucz} Atlas\,9 model atmosphere of a cool star (HD\,21749: $T_{\mathrm{eff}}$\,=\,4562\,K, $\log g$\,=\,4.58\,dex, $[Fe/H]$\,=\,0.02\,dex). 
We define a quantity to show the isolation degree of each \ion{Fe}{ii} line similarly to \cite{boeche2016}. We define the \ion{Fe}{ii} lines as blended if the EW of each potential blend 
relative to the sum of the EW of the iron line plus the EW of the potential blend is higher than a threshold value.\footnote{The expression of the blend threshold we use is: 
\\ threshold =  blend / (blend + iron line)} The higher the threshold value, the less lines will be considered blended. We selected three indicative threshold values of 10\%, 20\%, and 50\%. 
The results are shown in Table~\ref{lines}. With this analysis, we find that using a 10\% threshold, 11 lines appear to be blended: 4508.28, 4520.22, 4576.33, 4731.45, 4923.93, 5197.57, 
5337.72, 5991.37, 6149.25, 6442.97, and 6516.08\,\AA{}. 

\begin{figure*}%
 \centering
 \begin{minipage}{0.21\textwidth}
  \includegraphics[width=4.5cm, height=4.0cm]{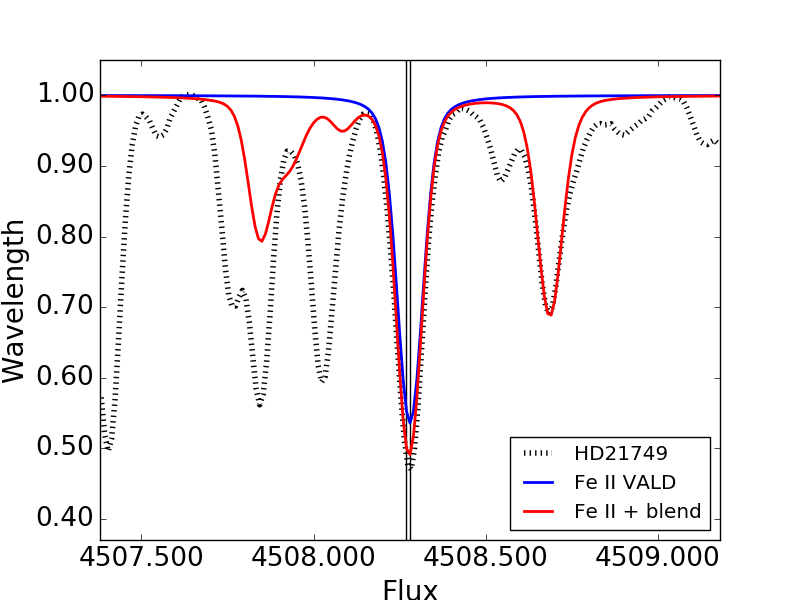} 
 \end{minipage}
\hspace{0.04\textwidth}%
 \begin{minipage}{0.21\textwidth}
  \includegraphics[width=4.5cm, height=4.0cm]{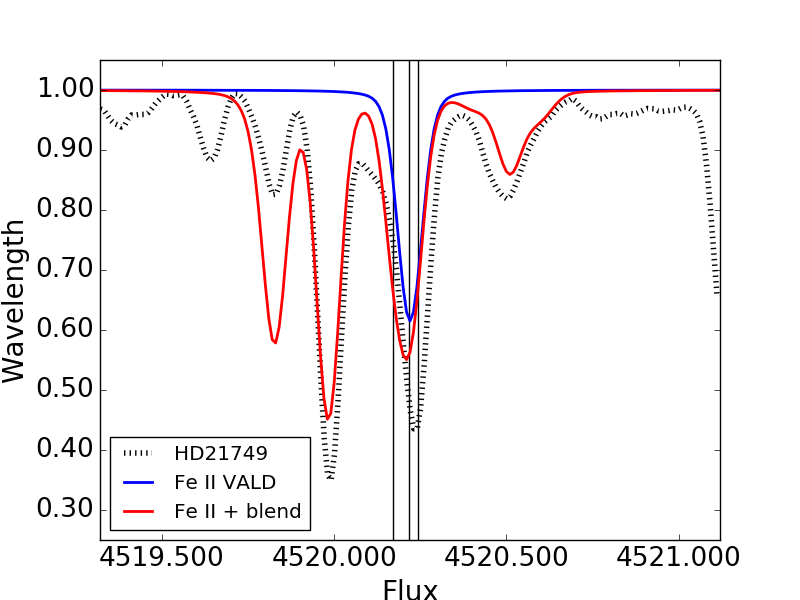} 
 \end{minipage}
\hspace{0.04\textwidth}%
 \begin{minipage}{0.21\textwidth}
  \includegraphics[width=4.5cm, height=4.0cm]{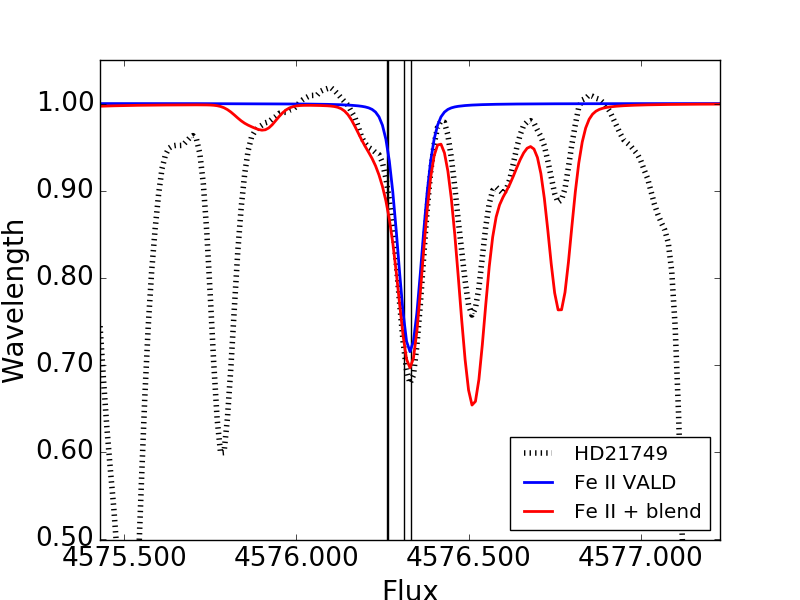} 
 \end{minipage}
\hspace{0.04\textwidth}%
 \begin{minipage}{0.21\textwidth}
  \includegraphics[width=4.5cm, height=4.0cm]{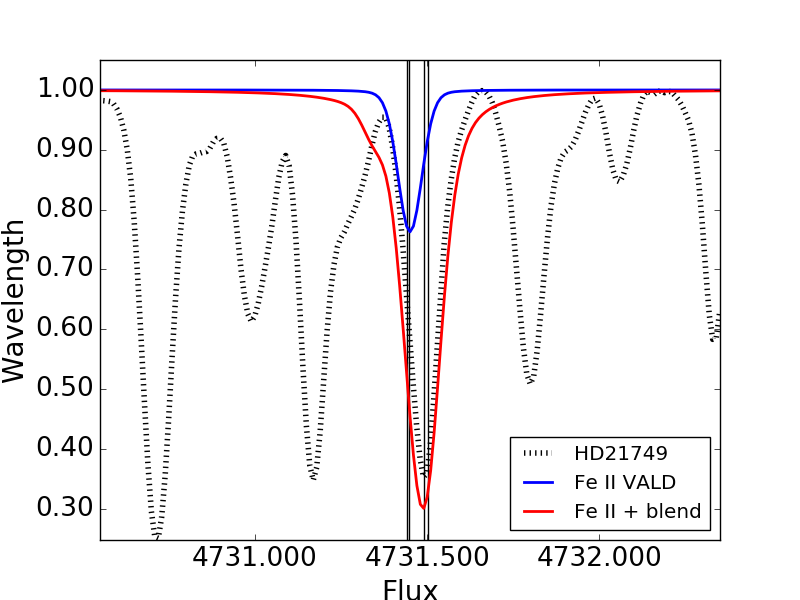} 
 \end{minipage}
 
 \centering
 \begin{minipage}{0.21\textwidth}
  \includegraphics[width=4.5cm, height=4.0cm]{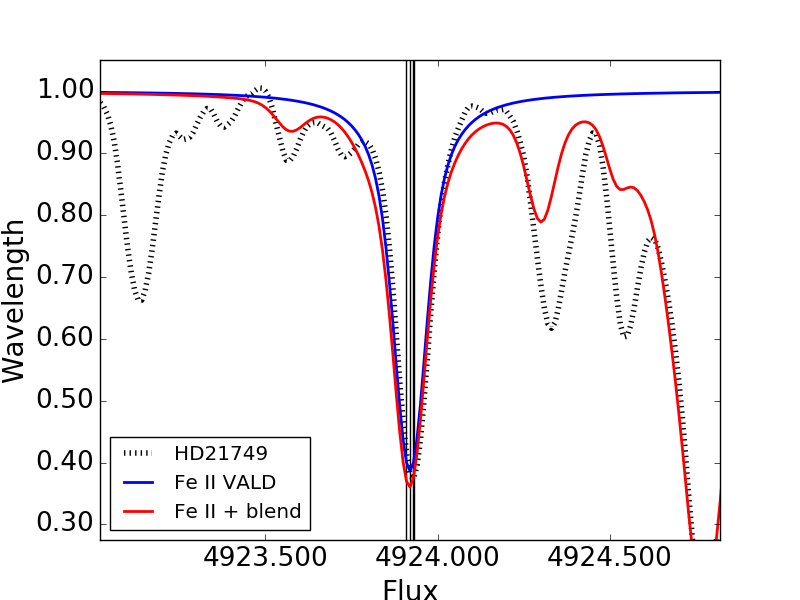} 
 \end{minipage}
\hspace{0.04\textwidth}%
 \begin{minipage}{0.21\textwidth}
  \includegraphics[width=4.5cm, height=4.0cm]{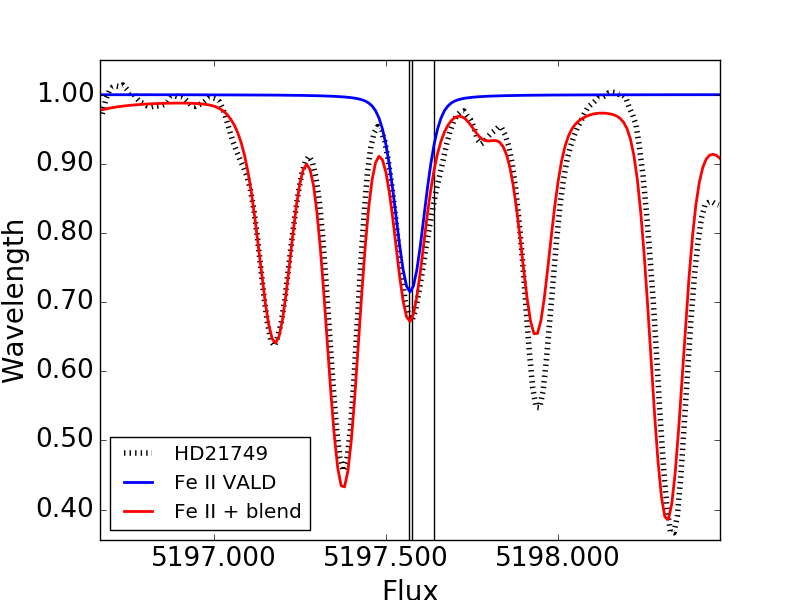} 
 \end{minipage}
\hspace{0.04\textwidth}%
 \begin{minipage}{0.21\textwidth}
  \includegraphics[width=4.5cm, height=4.0cm]{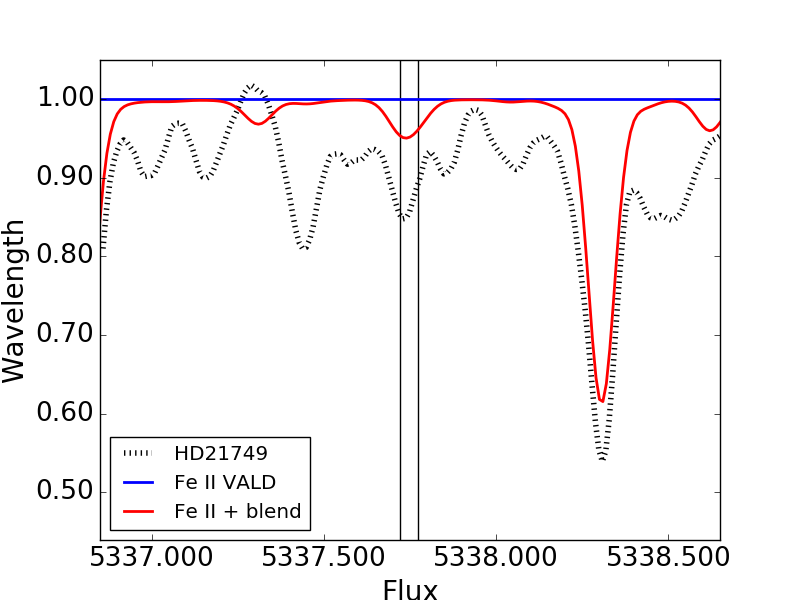} 
 \end{minipage}
\hspace{0.04\textwidth}%
 \begin{minipage}{0.21\textwidth}
  \includegraphics[width=4.5cm, height=4.0cm]{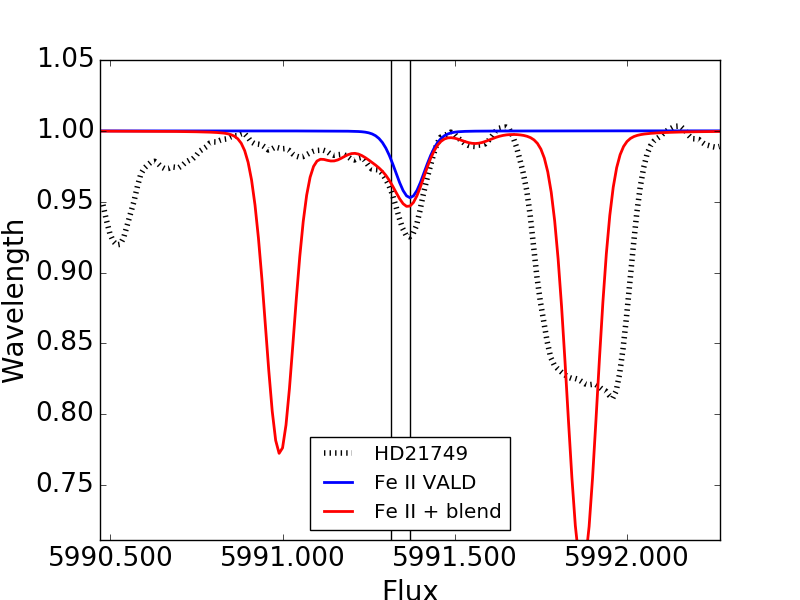} 
 \end{minipage}

 \centering
 \hspace{0.04\textwidth}%
 \begin{minipage}{0.21\textwidth}
  \includegraphics[width=4.5cm, height=4.0cm]{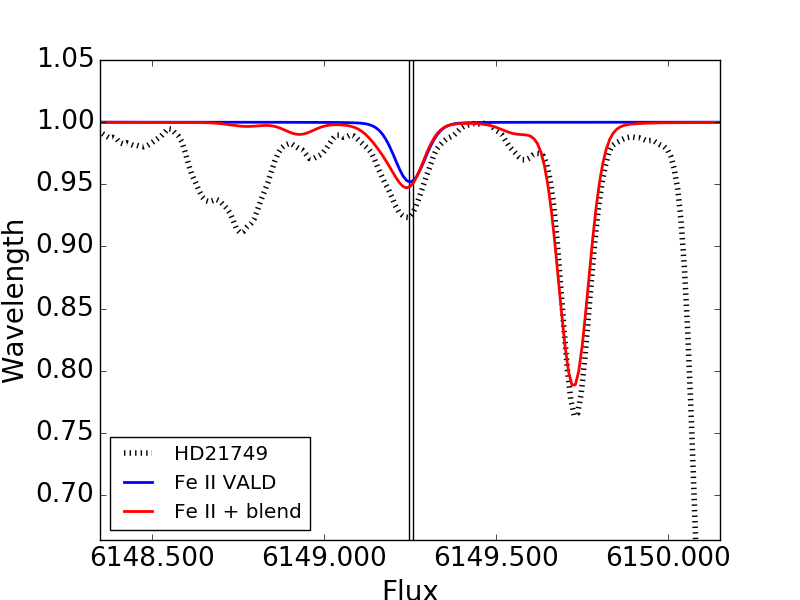} 
 \end{minipage}
 \hspace{0.04\textwidth}%
 \begin{minipage}{0.21\textwidth}
  \includegraphics[width=4.5cm, height=4.0cm]{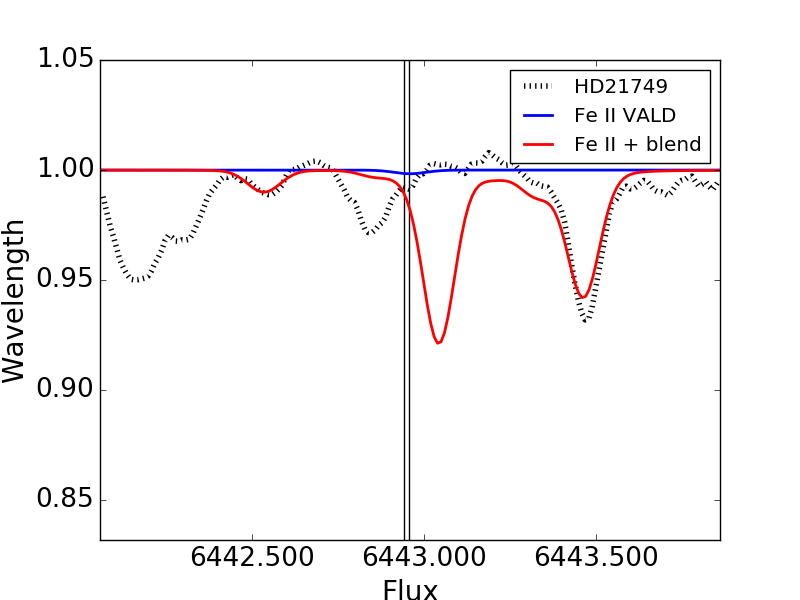} 
 \end{minipage}
\hspace{0.04\textwidth}%
 \begin{minipage}{0.21\textwidth}
  \includegraphics[width=4.5cm, height=4.0cm]{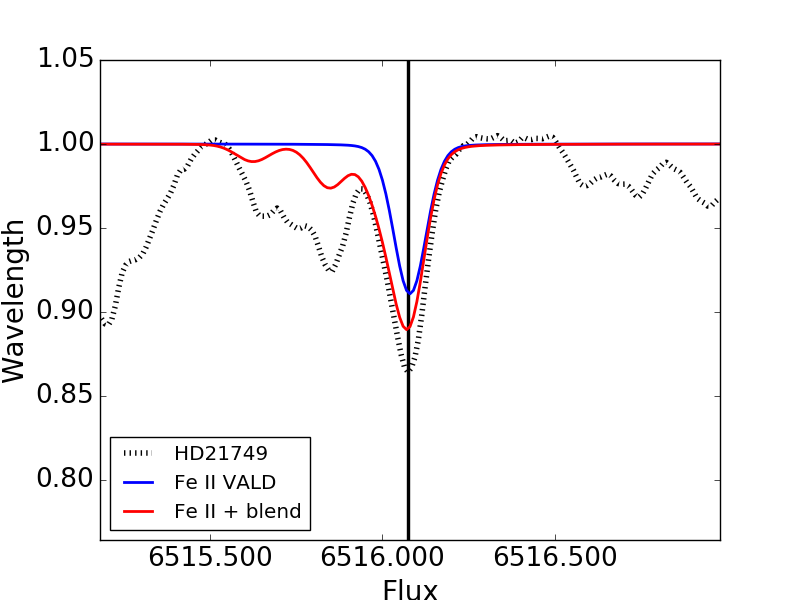} 
 \end{minipage}

 \caption{Synthetic \ion{Fe}{ii} lines in blue and the synthetic \ion{Fe}{ii} plus the blends in red. The HARPS spectrum of a K-type star (HD\,21749) is depicted (black dotted) for 
 comparison. The center of the \ion{Fe}{ii} line and the blends within the range of 0.20\AA{} are shown with the black solid vertical lines.}
 \label{synthetic_blends}
 \end{figure*}

In Fig.~\ref{synthetic_blends}, we use \texttt{MOOG} (\texttt{synth} driver) to synthesize the iron lines which appear blended, and their blends for the same model atmosphere. 
The lines which pass the 50\% blending threshold are 4731.45, 5337.72, and 6442.97\,\AA{}. The first appears visually most blended in Fig.~\ref{synthetic_blends} and the 
latter two are too weak (EW $\approx$ 0\,m\AA{}) for abundance analysis in this atmosphere. 

  \begin{table}
     \centering
      
     \caption{The blended \ion{Fe}{ii} lines based on VALD data are marked with a $\checkmark$ for each isolation threshold. The median and median absolute deviations (MAD) 
     are calculated for each \ion{Fe}{ii} line shown in Fig.~\ref{fe_abund_teff} for the two different damping options. N is the number of stars with $T_{\mathrm{eff}}$<\,5200\,K 
     from the HARPS sample. The lines with $\dag$ are considered 'bad' lines based on the criteria of Sect.~\ref{outliers_harps}. }
     \label{lines}
    $$  
        \scalebox{0.83}{
        \begin{tabular}{lcccccccc}
           \hline\hline
         \multirow{3}{*}{lines\,(\AA{})} & \multicolumn{3}{|c|}{Isolation threshold}      &  \multicolumn{2}{|c|}{Blackwell damping}  & \multicolumn{2}{|c|}{Barklem damping}  & \multirow{3}{*}{N}  \\
                                         & \multirow{2}{*}{10\%}  & \multirow{2}{*}{20\%} &  \multirow{2}{*}{50\%} &  median & MAD   & median & MAD   &                    \\                    
                                         &                        &                       &                        &  (dex)  & (dex) & (dex)  & (dex) &                    \\                    
         \hline
4508.28     & $\checkmark$ & $\checkmark$ &              & 0.04   & 0.04 & 0.06   & 0.05 & 130 \\
4520.22\dag & $\checkmark$ & $\checkmark$ &              & 0.27   & 0.10 & 0.28   & 0.10 & 130 \\
4576.34\dag & $\checkmark$ &              &              & 0.13   & 0.05 & 0.12   & 0.04 & 130 \\
4620.51     &              &              &              & --0.02 & 0.03 & --0.02 & 0.03 & 130 \\
4656.98     &              &              &              & -0.09  & 0.06 & --0.09 & 0.06 & 130 \\
4731.45\dag & $\checkmark$ & $\checkmark$ & $\checkmark$ & 0.43   & 0.10 & 0.45   & 0.10 & 130 \\
4923.93\dag & $\checkmark$ & $\checkmark$ &              & --0.04 & 0.22 & 0.01   & 0.23 & 129 \\
5197.57\dag & $\checkmark$ &              &              & 0.11   & 0.05 & 0.17   & 0.06 & 130 \\
5234.63     &              &              &              & --0.02 & 0.03 & 0.00   & 0.03 & 130 \\
5264.81     &              &              &              & 0.01   & 0.03 & --0.01 & 0.03 & 130 \\
5337.72\dag & $\checkmark$ & $\checkmark$ & $\checkmark$ & 0.42   & 0.30 & 0.44   & 0.30 & 86  \\
5414.07     &              &              &              & 0.00   & 0.03 & 0.00   & 0.03 & 129 \\
5991.38\dag & $\checkmark$ & $\checkmark$ &              & 0.13   & 0.06 & 0.13   & 0.07 & 127 \\
6149.25\dag & $\checkmark$ & $\checkmark$ &              & 0.14   & 0.06 & 0.14   & 0.05 & 130 \\
6247.56     &              &              &              & 0.06   & 0.03 & 0.08   & 0.03 & 130 \\
6442.97\dag & $\checkmark$ & $\checkmark$ & $\checkmark$ & 0.38   & 0.00 & 0.37   & 0.00 & 1   \\
6456.39\dag &              &              &              & 0.14   & 0.06 & 0.16   & 0.06 & 130 \\
6516.09\dag & $\checkmark$ & $\checkmark$ &              & 0.15   & 0.11 & 0.17   & 0.10 & 130 \\  
\hline

\end{tabular}}
    $$ 
  \end{table} 

Before we exclude any of these lines, we have to consider that because of inaccurate atomic data from VALD, their theoretical EW could either be over- or under-predicted and mislead us 
whether an iron line is blended or not. Moreover, the TS13 line list uses calibrated atomic data with respect to the Sun which means in case of a blended iron line in the solar 
spectrum, the calibrated $\log gf$ value is calculated from the EW of the total blend. If the blend has similar behavior as \ion{Fe}{ii}, the calibrated $\log gf$ can 
mitigate the effects of blending and probably will deliver a reliable abundance. Therefore, we need to combine other tests along with this analysis to confidently exclude the bad lines. 

\subsection{Searching for outliers using HARPS spectra}\label{outliers_harps}

We use a sample of 451 FGK-type dwarfs to estimate empirically which iron lines give outlying abundances that are too high to reconcile with the majority for each spectrum. 
The sample is part of the HARPS GTO planet search program \citep{mayor2003} with very high spectral quality, resolution of 110\,000, and 90\% of their spectra have signal-to-noise ratio 
higher than 200. The stellar parameters of this sample were derived by imposing excitation and ionization equilibria on weak iron lines with \texttt{MOOG}, using the 
\cite{kurucz} Atlas\,9, plane parallel, 1D static models in LTE. The same model atmospheres are used throughout this paper unless specified. These stars were firstly analyzed in terms of 
their parameters by \cite{sousa2008} and secondly by \cite{tsantaki13} with the same method but a shorter line list to correct for overestimations in the effective temperature 
for the cooler stars. As mentioned before, their effective temperatures and metallicities are in very good agreement with various spectroscopic and 
photometric works and thus, we consider these parameters very reliable to be used as reference.

\begin{figure}
  \centering
  \includegraphics[width=1.11\linewidth]{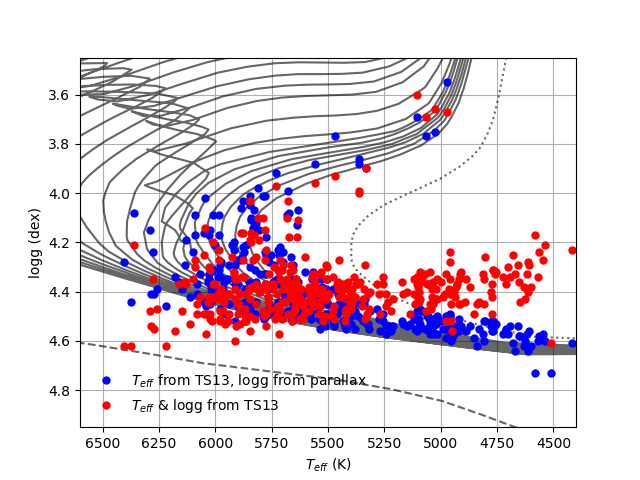} 
  \caption{The HR diagram for the 451 stars for the two set of parameters: the spectroscopic parameters of TS13 (red points), and spectroscopic $T_{\mathrm{eff}}$ and $\log g$ from \textit{Gaia} 
  parallaxes (blue points). The dotted line corresponds to 12.7\,Gyr isochrone of 0.035 metallicity, and the dashed line to 1\,Gyr isochrone of 0.0001 metallicity. The solid lines 
  correspond to isochrones between 1--12.7\,Gyr of solar metallicity.}
  \label{hr_ts13}
\end{figure}

Even though the effective temperatures agree very well with more model-independent methods, such as the IRFM, surface gravities on the other hand, appear flat in the 
Hertzsprung-Russell (HR) diagram for the cooler stars (Fig.~\ref{hr_ts13}). More reliable surface gravities are obtained from methods with less model 
dependence, such as asteroseismology or from dynamic mass and radius measurements in eclipsing binary systems. However, these measurements for dwarf stars are limited to a relatively small 
number. With the \textit{Gaia} mission \citep{Gaia2018}, we have access to parallaxes with unprecedented precision for millions of stars. In lack of the other direct $\log g$ estimates, 
trigonometric gravities are very useful to test our spectroscopic determinations. Trigonometric gravities are derived from the following expression: 

\begin{equation} 
 \log \frac{g}{g_{\odot}} = \log \frac{M}{M_{\odot}} + 4  \log \frac{T_{\mathrm{eff}}}{T_{\mathrm{eff}\odot}} + 0.4 (V + BC) + 2 \log \pi + 0.104 
\end{equation}
\\*
where M is the stellar mass, V the visual magnitude, BC the bolometric correction, and $\pi$ the parallax in mas. From the above expression, we derived the 
trigonometric $\log g$ for the 451 HARPS stars with the \textit{Gaia}\,DR2 parallaxes \citep{luri2018}, V magnitudes from the Hipparcos catalog 
\citep{Perryman1997}, bolometric correction based on \cite{flower96} and \cite{torres10}, solar magnitudes from \cite{bessell98}, and the $T_{\mathrm{eff}}$ of TS13. 
No correction for interstellar reddening is needed since all stars are less than 56\,pc in distance. Because of systematics in the 
\textit{Gaia}\,DR2 parallaxes, we add a conservative value of 0.03\,mas proposed by the \textit{Gaia} collaboration \citep{Lindegren2018}. Moreover, we 
increase the errors in parallaxes to consider the $\sim$30\% underestimation in uncertainties for bright stars \citep{luri2018, arenou2018}. Stellar 
masses are derived from the \texttt{PARAM\,1.3} tool\footnote{\texttt{PARAM\,1.3} tool: \href{http://stev.oapd.inaf.it/cgi-bin/param_1.3}{http://stev.oapd.inaf.it/cgi-bin/param\_1.3}} using the 
PARSEC theoretical isochrones from \cite{bressen2012} and a Bayesian estimation method \citep{dasilva06}. The trigonometric $\log g$ are also depicted in 
Fig.~\ref{hr_ts13} which follow the expected isochrones for the solar neighborhood.

First, we measure the EW of the complete TS13 line list for all spectra with \texttt{ARES}. 
Then, we use our reference spectroscopic $T_{\mathrm{eff}}$, $[Fe/H]$, and microturbulence from TS13, and the trigonometric gravities to derive the individual 
\ion{Fe}{i} and \ion{Fe}{ii} abundances for each star with \texttt{MOOG} (\texttt{abfind} driver) using a curve-of-growth approach. As mentioned before, the $\log gf$ values of the line 
list are calibrated to match the solar abundances ($\log A(Fe)_{\odot}$\,=\,7.47\,dex) while the collisional broadening parameters (van der Waals damping) 
are based on the \cite{unsold1955} approximation with an enhancement factor recommended by the Blackwell group\footnote{The enhancement factor, E, is given by the relation: 
E = 1 + 0.67$\cdot$EP, where EP is the excitation potential of the specific line.} (damping option 2 within MOOG for the 2017 version). 

Even though the strength of a weak line is dominated by the Doppler core rather than the Lorentzian wings, large uncertainties in the damping parameters affect the derived 
abundances even for weak lines \citep[see e.g.][]{ryan1998}. A different approach to determine damping parameters is described by the ABO theory \citep{Anstee1991, Barklem1998, Barklem2000} 
which uses cross sections to determine the individual damping values. The TS13 line list contains 115 out of 137 lines with damping based on the ABO theory from \cite{Barklem2000}. All of the 
\ion{Fe}{ii} lines have \cite{Barklem2000} damping data. We evaluate how the two different damping approaches affect the abundances of the \ion{Fe}{ii} lines. Before we do that, we recalibrate 
the $\log gf$ values for the damping of \cite{Barklem2000} using the Sun as reference with a solar spectrum from HARPS. The recalibration is essential because the initial TS13 line list was 
calibrated with the Blackwell damping.

\begin{figure}
   \centering
    \includegraphics[width=1.11\linewidth]{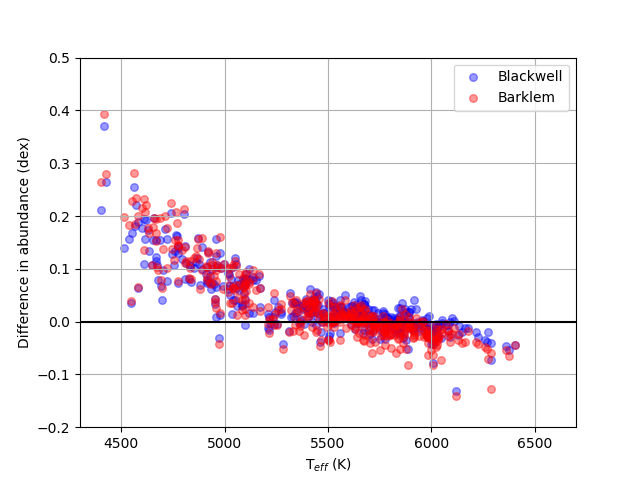}
   \caption{Differences in iron abundances (\ion{Fe}{ii} -- \ion{Fe}{i}) as a function of $T_{\mathrm{eff}}$ for the 451 stars for the two different damping options. Red points 
   represent the differences with Barklem damping values and the blue points the differences with Blackwell. The abundances are derived with the $T_{\mathrm{eff}}$, $[Fe/H]$, and microturbulence 
   from TS13, and the trigonometric $\log g$.}
   \label{fe_abund_teff}
\end{figure}

In Fig.~\ref{fe_abund_teff}, we plot the difference between \ion{Fe}{ii} -- \ion{Fe}{i} for the 451 stars using the different damping parameters. In the traditional LTE approach, the 
differences should be zero but in our case they are enhanced for the cooler stars ($\sim$0.3\,dex) and this is the typical ionization balance problem that is reported in the literature. We notice 
that there is no distinction from this plot which damping approach performs better as in both cases the differences in the iron abundances are equally large.

\begin{figure}
   \centering
   \includegraphics[width=1.1\linewidth]{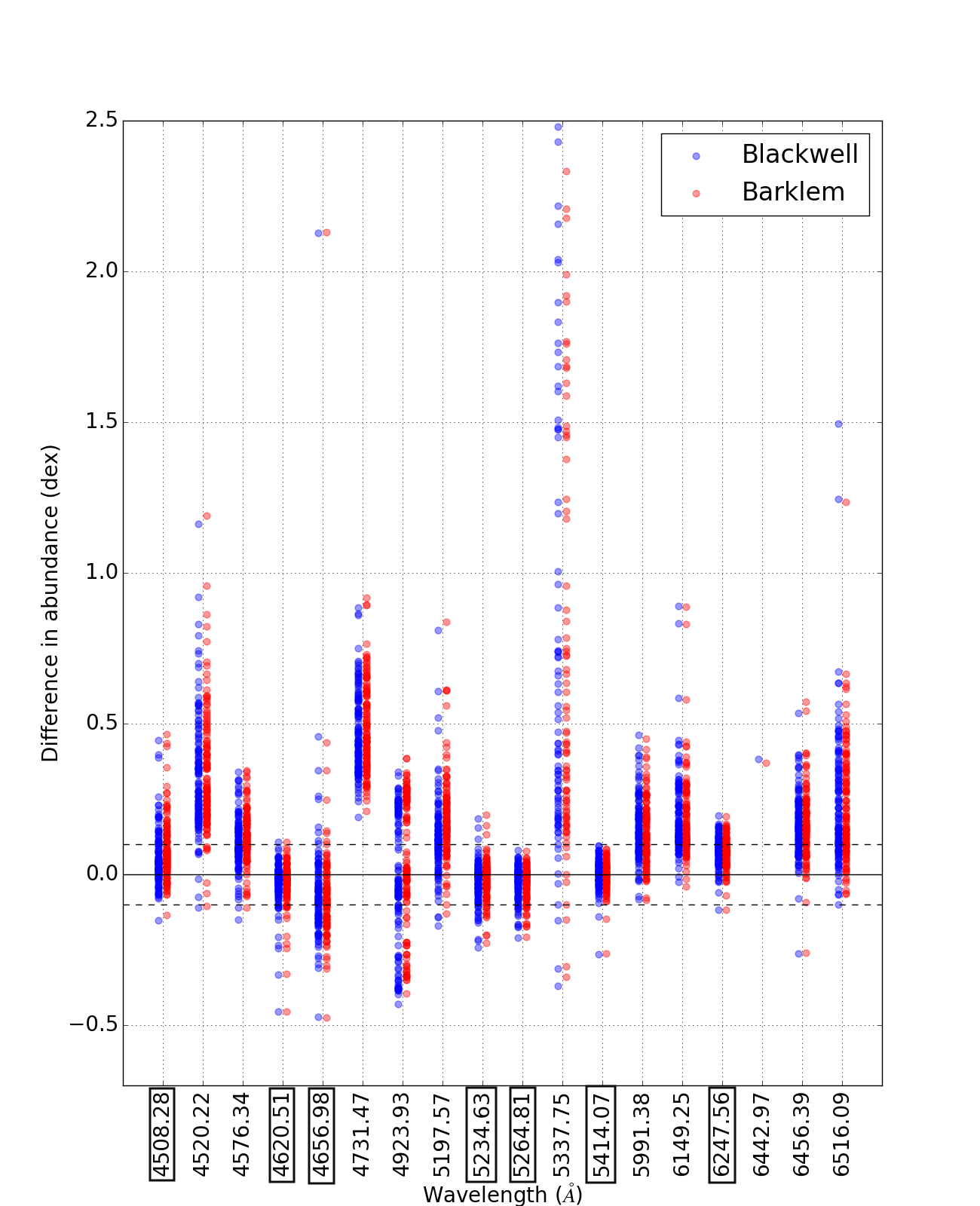}
   \caption{Difference in abundance (\ion{Fe}{ii} -- Fe) produced by each \ion{Fe}{ii} line for the cool stars of our sample. The dotted lines show the $\pm$0.10\,dex limit. 
   The different colors represent different damping options. The lines in squares are the ones selected by our criteria.}
   \label{feii_abund}
\end{figure}

Our goal is to select the \ion{Fe}{ii} lines which produce the same abundance as \ion{Fe}{i} (which is equal to the total iron abundance since it is measured only from \ion{Fe}{i} 
lines). In Fig.~\ref{feii_abund}, we present the differences of the overall iron abundance minus the abundance of each \ion{Fe}{ii} line for the cooler stars (134 stars) since the 
\ion{Fe}{ii} overabundances are more evident for $T_{\mathrm{eff}}$<\,5200\,K using the two different damping approaches. The results of all \ion{Fe}{ii} lines are also presented in 
Table~\ref{lines}. We exclude lines with median difference higher than 0.10\,dex and at the same time their mean absolute deviation is above 3\,$\sigma$ following a similar analysis as in 
\cite{Sousa2014}. For a cool star, a difference of 0.10\,dex in the \ion{Fe}{ii} -- \ion{Fe}{i} abundance can be produced by an underestimation of 0.15\,dex in $\log g$ which corresponds to 
the differences we roughly observe in Fig.~\ref{hr_ts13} between the spectroscopic and trigonometric $\log g$ and therefore can be a valid limit to exclude \ion{Fe}{ii} lines.
Both damping approaches give similar differences, within 0.02\,dex, but the highest reaches 0.06\,dex for the 5197.57\,\AA{}. In both cases the same lines are excluded. 

The remaining good lines are: 4508.28, 4620.51, 4656.98, 5234.63, 5264.81, 5414.07, 6247.56\,\AA{}. The line 6442.97\,\AA{} is too weak to appear in stars with 
$T_{\mathrm{eff}}$<\,5200\,K but we include it because for the hotter stars gives differences smaller than 0.10\,dex for the same analysis but for the 
rest of the sample). We note that for stars with $T_{\mathrm{eff}}$>\,5200\,K, all lines give differences below 0.10\,dex, except for 5337.72\,\AA{}. We confirm that 10 out of 11 lines 
from the previous analysis in Sect.~\ref{vald} are also classified as bad in this test, suggesting they are in fact blended. The reasons for the rest of the lines which give overabundances in 
this test could be related to bad EW measurements either because they are in regions difficult to normalize correctly or their atomic data are not accurate. To include these reasons, we 
adopt this \ion{Fe}{ii} line list as more robust compared to the one in Sect.~\ref{vald}. 

We also performed the same analysis but this time we used the MARCS \citep{Gustafsson2008} model atmospheres to check whether the iron abundances differ significantly depending 
on the choice of model atmospheres. To have meaningful results, we recalibrate the atomic data of our line list with a MARCS solar model atmosphere and derive the 
\ion{Fe}{ii} -- \ion{Fe}{i} abundances. We discover that the differences are almost identical to the ones derived with Kurucz models and subsequently the same lines were 
excluded. 

Summarizing, we exclude 55\% of the lines (10 out of 18) showing the difficulty to find good \ion{Fe}{ii} lines in the optical.

\subsection{Comparison with other line lists}

There are numerous iron line lists in the literature used for high resolution studies. In this Section, we check how two line lists perform in order to extend our number of good 
\ion{Fe}{ii} lines or obtain improved atomic data by comparing lines in common. The two line lists are: 12 \ion{Fe}{ii} lines from the \textit{Gaia}-ESO 'golden' list 
\citep{jofre2014}, and 120 \ion{Fe}{ii} lines from \cite{melendez2009}. We note that the line list of \cite{aleo2017} which was used to reduce the \ion{Fe}{ii} -- \ion{Fe}{i} 
differences for the Hyades cluster, contains 12 \ion{Fe}{ii} lines which are all selected from \cite{melendez2009}. 
The lines from the \textit{Gaia}-ESO 'golden' list were obtained by calculating the mean and standard deviation of all abundances for a sample of 34 FGKM stars and selected 
those lines that agreed within 2\,$\sigma$ with the average abundance  and had to be analyzed by at least three different research groups. 

We performed the analysis of Sect.~\ref{outliers_harps} for both line lists to show how many of these lines produce good abundance determinations. For this analysis, 
we used the two different damping options (when Barklem data were available), and we also did not find significant differences on the derived abundances. Therefore, for simplicity, we use only the Blackwell damping parameters for 
the rest of the paper since this option was used for the derivation of the initial parameters of the HARPS sample we compare in the following sections, and more importantly, since we showed 
that the damping does not affect significantly the results of this analysis. We select the lines which fulfill the criteria of Sect.~\ref{outliers_harps} resulting in only two good lines from 
\cite{jofre2014} and 15 from \cite{melendez2009}. The two lines from the \textit{Gaia}-ESO line list (5264.81, 5414.07\,\AA{}) are in common with the good lines of 
\cite{melendez2009} and show the same median abundances but with the atomic data of TS13, their median values are smaller, and therefore, we keep the latter atomic data. 
There are six lines from \cite{melendez2009} which are not in the TS13 list and are now included as new good lines. 

From the lines in common between TS13 and \cite{melendez2009}, we select the atomic data of those which present the lowest median and dispersion values. For example, the line 4576.34\,\AA{} 
with the solar calibrated $\log gf$ from TS13 gives difference in the abundances of 0.13\,dex, yet with the atomic data of \cite{melendez2009} the difference is reduced to 0.07\,dex. 
Inversely, the 4508.29\,\AA{} which was excluded by \cite{aleo2017} as blended according to their criteria, does not give overabundances with the atomic data of TS13. 
This test demonstrates that the atomic data play an important role, as important as the correct EW measurement itself, on the abundance analysis. The final line list contains 14 lines from 
both \cite{melendez2009} and TS13 and is described in Table~\ref{lines_new}. We remind the reader that the atomic data for the TS13 line list are derived after solar calibration with the 
damping Blackwell approximation. 

  \begin{table}
     \centering
      
     \caption{ The final \ion{Fe}{ii} lines from the analysis of all line lists with the excitation potential (EP) and the optimal atomic data taken from the reference. }
     \label{lines_new}
        \begin{tabular}{lccc}
           \hline\hline
         lines\,($\AA{}$) & EP (ev) & $\log gf$ & Reference \\
           \hline
4369.41 & 2.78 & --3.650 & \cite{melendez2009} \\
4508.28 & 2.86 & --2.405 & TS13 \\
4522.63 & 2.84 & --2.250 & \cite{melendez2009} \\
4576.34 & 2.84 & --2.950 & \cite{melendez2009} \\
4582.84 & 2.84 & --3.180 & \cite{melendez2009} \\
4620.51 & 2.83 & --3.236 & TS13 \\
4656.98 & 2.89 & --3.679 & TS13 \\
4666.76 & 2.83 & --3.280 & \cite{melendez2009} \\
5234.63 & 3.22 & --2.237 & TS13 \\
5264.81 & 3.23 & --3.093 & TS13 \\
5414.07 & 3.22 & --3.571 & TS13 \\
6239.95 & 3.89 & --3.410 & \cite{melendez2009} \\
6247.56 & 3.89 & --2.300 & \cite{melendez2009} \\
6442.97 & 5.55 & --2.400 & TS13 \\
\hline
        \end{tabular}
  \end{table} 

\section{Ionization balance of iron for the 451 HARPS stars}

Now, we perform the inverse exercise to show that in fact the clean \ion{Fe}{ii} list we compiled previously provides more reliable surface gravities. We derive the atmospheric parameters 
for the sample using the spectral analysis tool \texttt{FASMA}\footnote{\texttt{FASMA} tool: \href{http://www.iastro.pt/fasma}{http://www.iastro.pt/fasma}} \citep{Andreasen2017}. \texttt{FASMA} is wrapped around 
\texttt{ARES} and \texttt{MOOG}, and includes the model interpolation and minimization processes. The parameters are derived based on the same principles as in TS13, i.e. imposing excitation 
balance (the slope of iron abundance and excitation potential should be lower than 0.001) and ionization balance (the difference between the average abundances of \ion{Fe}{i} and 
\ion{Fe}{ii} should be less than 0.01\,dex) in a fully automatic way. These convergence criteria are defined in \cite{Andreasen2017}. We apply a 3\,$\sigma$ clipping to remove outliers.

\begin{figure}
  \centering
   \includegraphics[width=1.1\linewidth]{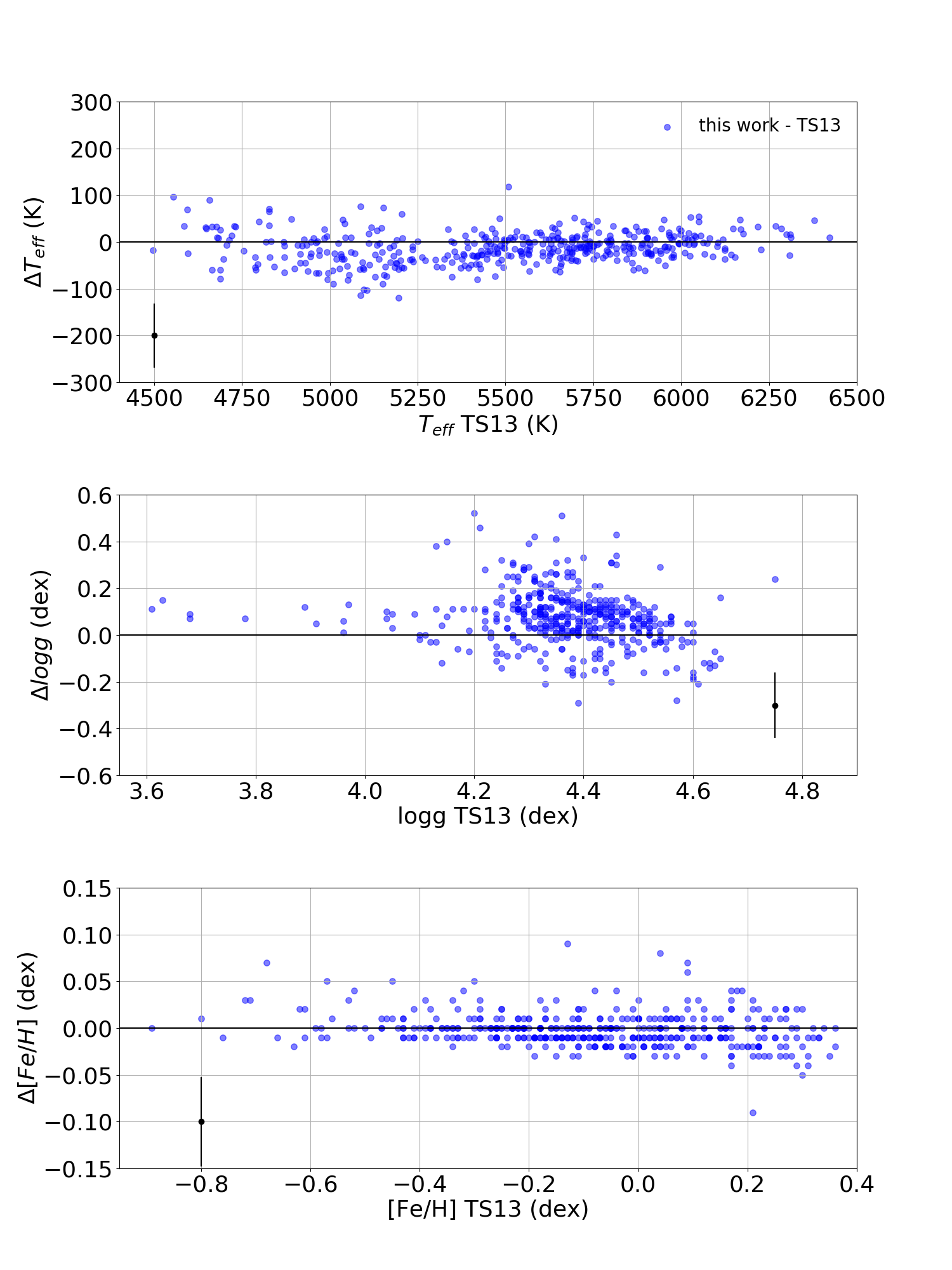}    
  \caption{Differences in effective temperature (upper plot), surface gravity (middle plot) and metallicity (bottom plot) between this work and TS13. The average error bars are plotted as black points.}
  \label{451_params}
  \end{figure}

\begin{figure}
  \centering
   \includegraphics[width=1.1\linewidth]{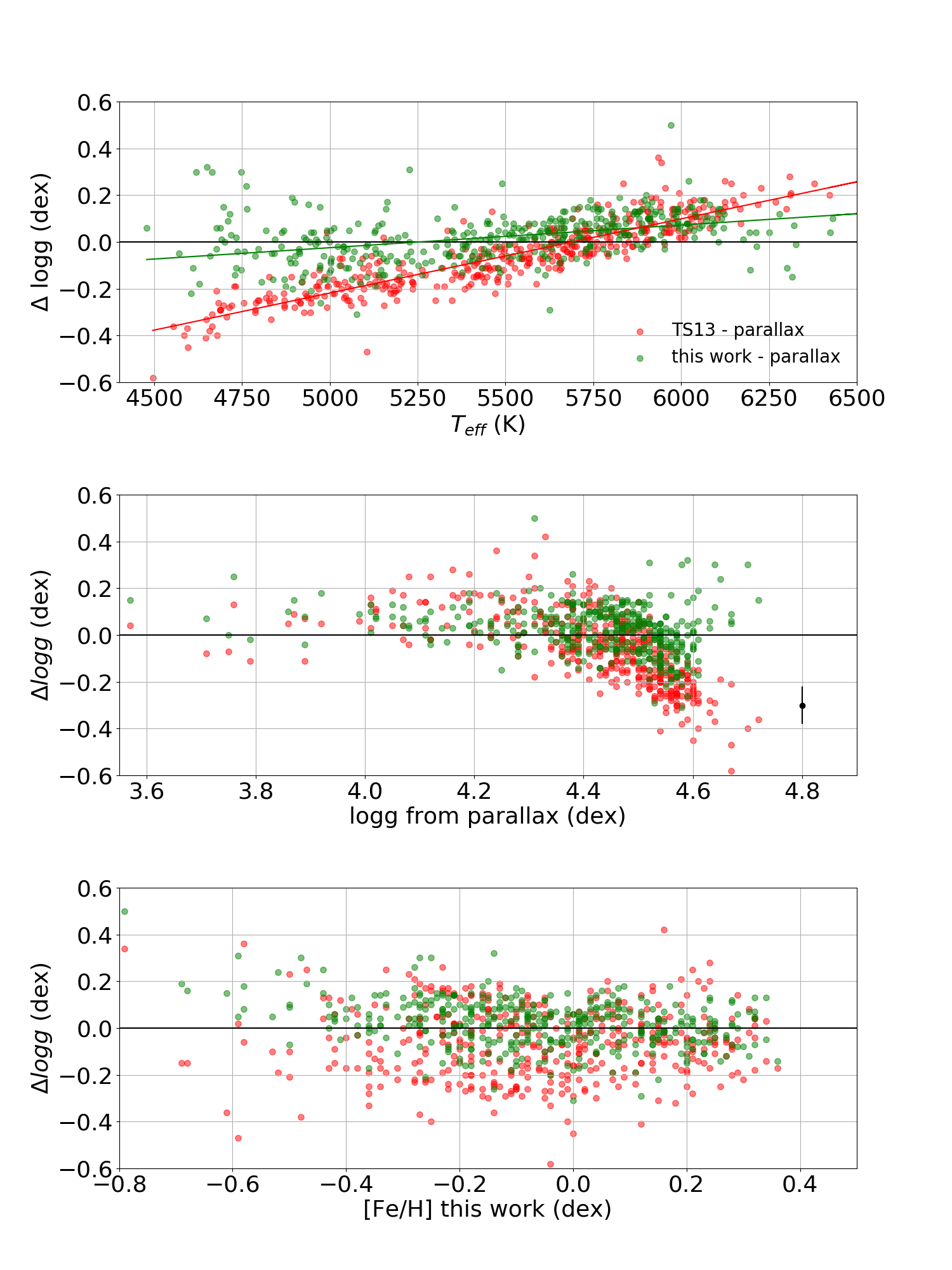}  
  \caption{Differences in $\log g$ of this work with trigonometric (green points), and of TS13 with trigonometric (red points), for the different atmospheric 
  parameters: $\Delta\log g$ -- $T_{\mathrm{eff}}$ (upper plot), $\Delta\log g$ -- trigonometric $\log g$ (middle plot), $\Delta\log g$ -- $[Fe/H]$ (bottom plot). The colored lines in the 
  upper plot represent a linear fit of the two samples and the black point in the middle plot the average error.}
  \label{451_logg}
  \end{figure}

In Fig.~\ref{451_params}, we show our derived parameters compared with TS13. The mean difference and standard deviation between \texttt{FASMA} and TS13 for $T_{\mathrm{eff}}$ is --11 and 34\,K, 
for $\log g$ is 0.08 and 0.13\,dex and for $[Fe/H]$ is 0.00 and 0.02\,dex. We show that the scale of the effective temperature and metallicity of this work is in very good agreement 
with TS13 which we expect because the \ion{Fe}{i} lines did not change. 

In Fig.~\ref{451_logg} (middle plot), we plot the comparison between our spectroscopic surface gravities and the 
trigonometric. The mean differences and standard deviations between the $\log g$ of this work and the trigonometric is 0.02 and 0.10\,dex, and between the $\log g$ of TS13 and trigonometric 
is --0.06 and 0.13\,dex. Even though the agreement of our spectroscopic $\log g$ with the trigonometric is very good, we notice an overestimation of our $\log g$ for values higher than $\sim$4.60\,dex. 
Nevertheless, we see a clear improvement in the $\Delta\log g$ -- $T_{\mathrm{eff}}$ relation in this work 
(upper plot of Fig.~\ref{451_logg}) and our dispersion values are smaller. These results indicate that the criteria of the line selection were efficient enough because that linear dependence 
between $\Delta\log g$ and $T_{\mathrm{eff}}$ has diminished significantly with a slope of practically zero (9.7$\cdot$10$^{-5}$\,dex\,K$^{-1}$). 
Finally, there is no general correlation between $\Delta\log g$ and $[Fe/H]$ (bottom plot of Fig.~\ref{451_logg}) but there is an overestimation of spectroscopic 
$\log g$ for the most metal-poor stars of this sample ($[Fe/H]<$\,--0.4\,dex). These are the same stars which produce the overestimations in the middle plot with $\log g >$\,4.60\,dex.  
It is difficult to explain why metal-poor stars do not still have well constrained gravities. These differences cannot be attributed to non-LTE effects since these effects would cause an 
underestimation of $\log g$ and not an overestimation we observe here (see discussion below). 

\begin{figure}
   \centering
    \includegraphics[width=1.15\linewidth]{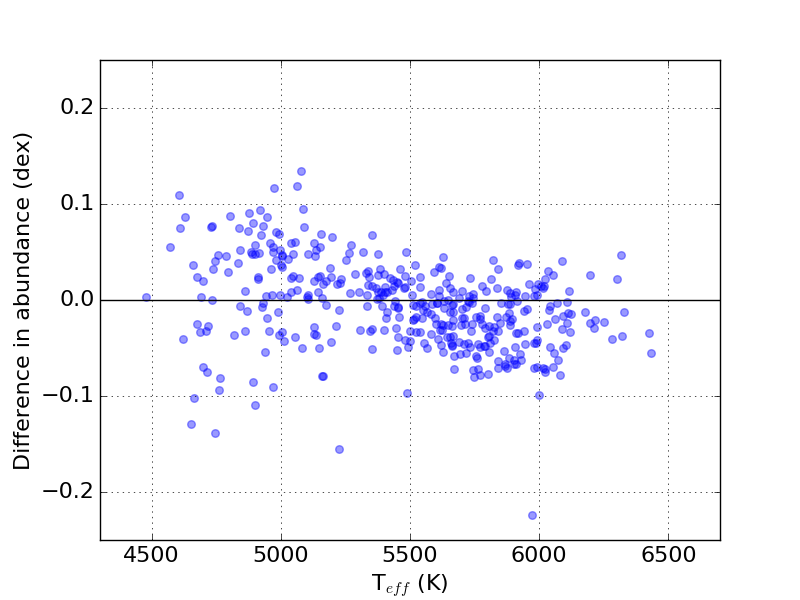} 
   \caption{Difference between \ion{Fe}{ii} -- \ion{Fe}{i} abundances for the stars of our sample with the clean \ion{Fe}{ii} list $T_{\mathrm{eff}}$.}
   \label{fe_abund_teff_clean}
\end{figure}

Using the spectroscopic $T_{\mathrm{eff}}$ and $[Fe/H]$, and the trigonometric gravity, we calculate the new \ion{Fe}{ii} and \ion{Fe}{i} abundances for our sample similarly as in 
Sect.~\ref{outliers_harps}. In Fig.~\ref{fe_abund_teff_clean}, we plot the difference in \ion{Fe}{ii} -- \ion{Fe}{i} with $T_{\mathrm{eff}}$ for the whole sample of 451 stars. 

This plot shows that ionization balance is on average fulfilled for the lines we consider clean after the blending analysis. The \ion{Fe}{ii} -- 
\ion{Fe}{i} mean difference is --0.01\,dex and the standard deviation is 0.04\,dex, and thus, we can say that the ionization balance is much better satisfied. 
The highest dispersion appears for the cooler stars. 
The remaining differences can be attributed to the effects we did not consider in this work. For instance, \cite{lind2012} show that the \ion{Fe}{i} abundances are underestimated under the LTE 
assumption and the amount of non-LTE corrections depends on stellar parameters. The highest correction for non-LTE effects in \ion{Fe}{i} abundance is expected for hot, metal-poor giants and 
for the parameters of our sample it should be less than $\sim$0.02--0.03\,dex (for the case of a hot dwarf, e.g., $T_{\mathrm{eff}}$\,=\,6000\,K, $\log g$\,=\,4.0\,dex, [Fe/H]\,=\,0.1\,dex), 
estimated from their figure 4. There are no hot metal-poor stars in our sample. \cite{bensby2014} faced the ionization balance problem in their study of a large sample of F and G dwarf 
stars in the solar neighborhood. The authors pointed out that stars with $\log g>$\,4.20\,dex and $T_{\mathrm{eff}}<$\,5600\,K showed higher \ion{Fe}{ii} abundances over \ion{Fe}{i} and 
non-LTE effects could not explain the large differences between spectroscopic and trigonometric gravities. However, they provide empirical calibrations for the coolest stars to correct for the 
discrepancies in the $\log g$ determinations. 

\cite{jofre2014} also find discrepancies between \ion{Fe}{ii} and \ion{Fe}{i} abundances for their sample which are higher for the K-dwarfs (their table 3). 
The authors also quantified the departures from LTE for their sample to be usually less than 0.03\,dex for the FGK dwarfs and concluded that non-LTE effects are too small to justify the ionization 
balance problem. 

The differences of Fig.~\ref{fe_abund_teff_clean} are small enough to be caused by other effects, such as activity, but also inaccuracies at this level of our reference stellar parameters 
should not be excluded.

\section{Ionization balance of titanium for the 451 HARPS stars}
 
In the previous Section, we showed that a careful selection \ion{Fe}{ii} lines can improve the trends of iron abundances with effective temperature and provide $\log g$ in agreement with 
astrometry. However, other species could be more suitable for the surface gravity determinations. We search for species with enough neutral and ionized lines in the optical in two line 
lists: \textit{i)} from \cite{neves09}, and \textit{ii)} from the joint list of \cite{lawler2013} and \cite{wood2013}, hereafter LW13. 

\subsection{The Ti line list of Neves et al. (2009)}

The element with most numerous ionized lines after iron is Ti in this line list (Sc and Cr follow after) and contains 30 \ion{Ti}{i} and 8 \ion{Ti}{ii} 
lines. Our goal is to check whether the surface gravities derived from the ionization equilibrium of Ti are more precise and accurate than from Fe. 

\begin{figure}
  \centering
   \includegraphics[width=1.1\linewidth]{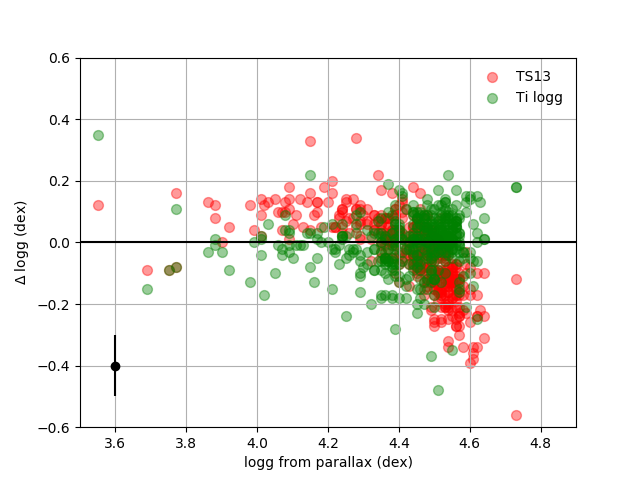} \\   
   \includegraphics[width=1.1\linewidth]{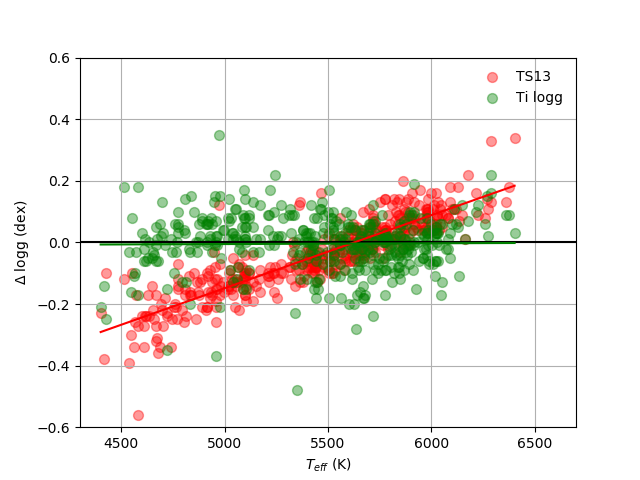}  
  \caption{Differences in $\log g$ of the TS13 (red points) with the trigonometric and from Ti of this work with the line list of Neves et al. (2009) (green points) with the trigonometric as 
  a function of $\log g$ (upper plot) and $T_{\mathrm{eff}}$ (bottom plot). The colored lines in the bottom plot represent the linear fits of the two samples. The slope of the Ti (green) 
  line is  2.4$\cdot$10$^{-6}$\,dex\,K$^{-1}$.}
  \label{ti_logg}
  \end{figure}

The atomic data of this list were also derived after solar calibration and the damping parameters used for their analysis is based on the Blackwell approximation which we also used in this 
Section. The EW of \ion{Ti}{i} and \ion{Ti}{ii} lines are measured with \texttt{ARES} for the whole sample of Sect.~\ref{outliers_harps}. We modify \texttt{FASMA} to obtain surface gravities 
from the ionization equilibrium of Ti having the other stellar parameters fixed to the values of TS13. The mean difference and standard deviation between $\log g$ derived from Ti and the 
trigonometric is 0.00 and 0.09\,dex and are smaller than the derived from the Fe ionization equilibrium (upper plot of Fig.~\ref{ti_logg}). Additionally, the slope of $\Delta \log g$ and 
$T_{\mathrm{eff}}$ is $\sim$0 (bottom plot of Fig.~\ref{ti_logg}). 
We notice that there is an overestimation of the spectroscopic gravities for the hottest stars of the sample ($T_{\mathrm{eff}}>$\,6100\,K). 

\subsection{The Ti line list of LW13}

The line list of LW13 provides $log gf$ data from the measurements of the branching fractions in solar FTS and echelle spectra. From the 948 \ion{Ti}{i} lines of \cite{lawler2013} 
and the 364 \ion{Ti}{ii} lines of \cite{wood2013}, the authors defined 128 and 31 lines respectively in the optical suitable for the solar abundance determination. However, it is not guaranteed 
this line list will work for cooler atmospheres because severe blending occurs in lower temperatures. In fact, we used the complete 'solar' LW13 line list to derive surface gravities but with 
unrealistic results. A fast and efficient way to select the best lines is to perform a similar test as in Sect.~\ref{outliers_harps} for the HARPS sample, by measuring the EW of \ion{Ti}{i} and 
\ion{Ti}{ii} lines with \texttt{ARES} and deriving their abundances. We select the lines with \ion{Ti}{ii} -- \ion{Ti}{i} difference less than 0.10\,dex and dispersion lower than 3\,$\sigma$. 
The clean line list of LW13 now contains 59 \ion{Ti}{i} and 11 \ion{Ti}{ii}. The damping parameters for this line list are based on the Blackwell approximation.

\begin{figure}
  \centering
   \includegraphics[width=1.1\linewidth]{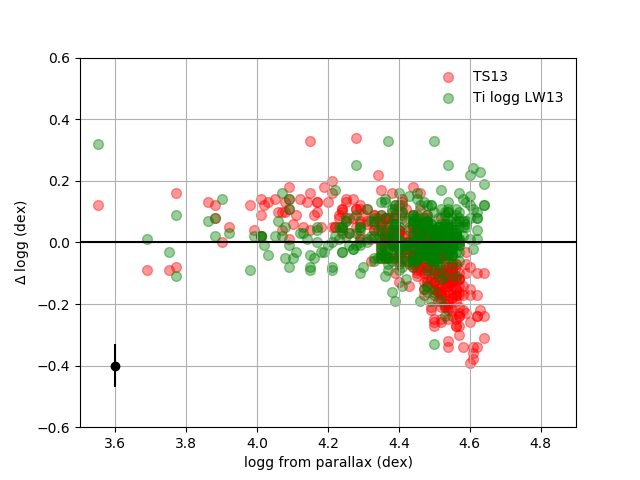} \\   
   \includegraphics[width=1.1\linewidth]{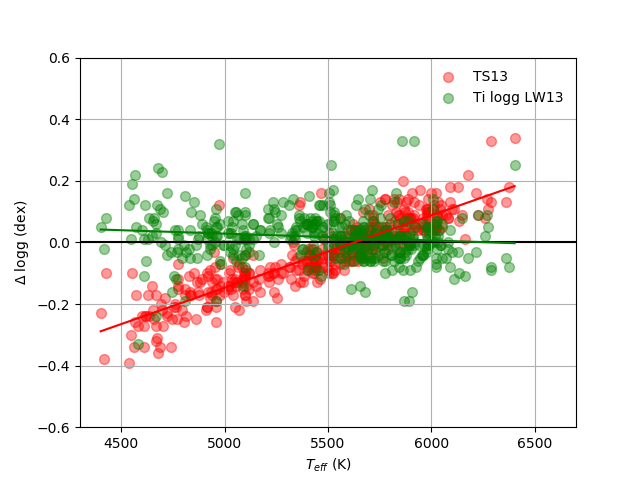}  
  \caption{ Differences in $\log g$ of the TS13 (red points) with the trigonometric and from Ti of this work with the clean line list of LW13 (green points) with the trigonometric as a 
  function of $\log g$ (upper plot) and $T_{\mathrm{eff}}$ (bottom plot). The colored lines in the bottom plot represent the linear fits of the two samples. The slope of the Ti (green) 
  line is --2.2$\cdot$10$^{-5}$\,dex\,K$^{-1}$.}
  \label{ti_logg_lw13}
  \end{figure}

\begin{table}
 \centering
   \label{ti_lw13}
   \caption{The final Ti lines we extracted from the LW13 line list.}
   \begin{tabular}{lccc}
   \hline\hline
   lines\,($\AA{}$) & EP (eV) & $\log gf$ & Element \\
   \hline
4512.73 & 0.84 & --0.40 & \ion{Ti}{i} \\ 
4518.69 & 1.43 & --1.04 & \ion{Ti}{i} \\ 
4623.10 & 1.74 &   0.16 & \ion{Ti}{i} \\ 
4639.36 & 1.74 & --0.05 & \ion{Ti}{i} \\ 
4645.19 & 1.73 & --0.51 & \ion{Ti}{i} \\ 
4650.01 & 1.74 & --0.64 & \ion{Ti}{i} \\ 
4656.04 & 1.75 & --0.99 & \ion{Ti}{i} \\ 
4731.16 & 2.17 & --0.43 & \ion{Ti}{i} \\ 
4733.42 & 2.16 & --0.66 & \ion{Ti}{i} \\ 
4742.11 & 2.15 & --0.94 & \ion{Ti}{i} \\ 
4747.67 & 2.25 & --0.83 & \ion{Ti}{i} \\ 
4778.26 & 2.24 & --0.35 & \ion{Ti}{i} \\ 
4820.41 & 1.50 & --0.38 & \ion{Ti}{i} \\ 
4926.15 & 0.82 & --2.09 & \ion{Ti}{i} \\ 
4997.10 & 0.00 & --2.07 & \ion{Ti}{i} \\ 
5020.03 & 0.84 & --0.33 & \ion{Ti}{i} \\ 
5022.87 & 0.83 & --0.33 & \ion{Ti}{i} \\ 
5024.84 & 0.82 & --0.53 & \ion{Ti}{i} \\ 
5064.65 & 0.05 & --0.94 & \ion{Ti}{i} \\ 
5109.43 & 1.44 & --1.54 & \ion{Ti}{i} \\ 
5145.46 & 1.46 & --0.54 & \ion{Ti}{i} \\ 
5147.48 & 0.00 & --1.94 & \ion{Ti}{i} \\ 
5152.18 & 0.02 & --1.95 & \ion{Ti}{i} \\ 
5210.38 & 0.05 & --0.82 & \ion{Ti}{i} \\ 
5212.99 & 2.23 & --1.16 & \ion{Ti}{i} \\ 
5230.97 & 2.24 & --1.19 & \ion{Ti}{i} \\ 
5295.78 & 1.07 & --1.59 & \ion{Ti}{i} \\ 
5366.64 & 0.82 & --2.46 & \ion{Ti}{i} \\ 
5384.63 & 0.83 & --2.77 & \ion{Ti}{i} \\ 
5389.99 & 1.87 & --1.10 & \ion{Ti}{i} \\ 
5448.91 & 2.33 & --1.25 & \ion{Ti}{i} \\ 
5449.15 & 1.44 & --1.87 & \ion{Ti}{i} \\ 
5453.64 & 1.44 & --1.60 & \ion{Ti}{i} \\ 
5471.19 & 1.44 & --1.42 & \ion{Ti}{i} \\ 
5474.22 & 1.46 & --1.23 & \ion{Ti}{i} \\ 
5474.45 & 2.34 & --0.95 & \ion{Ti}{i} \\ 
5490.15 & 1.46 & --0.84 & \ion{Ti}{i} \\ 
5503.90 & 2.58 & --0.05 & \ion{Ti}{i} \\ 
5512.52 & 1.46 & --0.40 & \ion{Ti}{i} \\ 
5514.34 & 1.43 & --0.66 & \ion{Ti}{i} \\ 
5514.53 & 1.44 & --0.50 & \ion{Ti}{i} \\ 
5739.47 & 2.25 & --0.61 & \ion{Ti}{i} \\ 
5823.69 & 2.27 & --1.01 & \ion{Ti}{i} \\ 
5880.27 & 1.05 & --2.00 & \ion{Ti}{i} \\ 
5899.29 & 1.05 & --1.10 & \ion{Ti}{i} \\ 
5999.66 & 2.24 & --0.72 & \ion{Ti}{i} \\ 
6017.55 & 2.33 & --1.69 & \ion{Ti}{i} \\ 
6091.17 & 2.27 & --0.32 & \ion{Ti}{i} \\ 
6092.79 & 1.89 & --1.38 & \ion{Ti}{i} \\ 
6121.00 & 1.88 & --1.42 & \ion{Ti}{i} \\ 
6146.21 & 1.87 & --1.48 & \ion{Ti}{i} \\ 
6149.73 & 2.16 & --1.44 & \ion{Ti}{i} \\ 
6258.10 & 1.44 & --0.39 & \ion{Ti}{i} \\ 
6258.71 & 1.46 & --0.28 & \ion{Ti}{i} \\ 
6266.01 & 1.75 & --1.95 & \ion{Ti}{i} \\ 
6268.53 & 1.43 & --2.26 & \ion{Ti}{i} \\ 
6336.10 & 1.44 & --1.69 & \ion{Ti}{i} \\ 
6419.09 & 2.17 & --1.53 & \ion{Ti}{i} \\ 
6745.54 & 2.24 & --1.23 & \ion{Ti}{i} \\ 
   \hline
   \end{tabular} 
\end{table} 

\begin{table}
\centering
\contcaption{ }
  \begin{tabular}{lccc}
  \hline\hline
  lines\,($\AA{}$) & EP (eV) & $\log gf$ & Element \\
  \hline
4330.24 & 2.05 & --1.64 & \ion{Ti}{ii} \\ 
4421.94 & 2.06 & --1.64 & \ion{Ti}{ii} \\ 
4443.80 & 1.08 & --0.71 & \ion{Ti}{ii} \\ 
4450.48 & 1.08 & --1.52 & \ion{Ti}{ii} \\ 
4493.52 & 1.08 & --2.78 & \ion{Ti}{ii} \\ 
4518.33 & 1.08 & --2.56 & \ion{Ti}{ii} \\ 
4583.41 & 1.16 & --2.84 & \ion{Ti}{ii} \\ 
4657.20 & 1.24 & --2.29 & \ion{Ti}{ii} \\ 
4708.66 & 1.24 & --2.35 & \ion{Ti}{ii} \\ 
4719.51 & 1.24 & --3.32 & \ion{Ti}{ii} \\ 
4874.01 & 3.09 & --0.86 & \ion{Ti}{ii} \\
  \hline
  \end{tabular}
\end{table} 

We obtain surface gravities from the ionization equilibrium of Ti with \texttt{FASMA}. 
The results are depicted in Fig.~\ref{ti_logg_lw13} where the mean difference and standard deviation between $\log g$ derived from Ti and the trigonometric is 0.02 and 0.08\,dex respectively, 
and the slope of $\Delta \log g$ and $T_{\mathrm{eff}}$ is also $\sim$0. 
There is a small overestimation of spectroscopic $\log g$ in this case for the cooler stars and not for the hottest ones as in Fig.~\ref{ti_logg}. The differences in these figures are on the 
opposite direction with $T_{\mathrm{eff}}$ and suggest that non-LTE effects should not be cause because they would affect the abundances the same way but probably we can attribute 
them on errors on the line selection. However, since these over-, under-estimations are small with both line lists, we consider already the results reliable enough.

Both Ti line lists in this Section give smaller standard deviations than Fe and flattest slopes with $T_{\mathrm{eff}}$. Therefore, Ti is a better indicator to probe surface 
gravity than Fe, once the other stellar parameters are defined. \cite{Bergemann2011} studied the non-LTE effects of Ti in the atmospheres of very metal-poor stars ($[Fe/H]<$--1.28\,dex) and 
suggested that they should not be ignored in the case of giant stars at low metallicities. Even though these parameters are outside our parameter space, one should be careful when applying this 
method without the necessary corrections to these stars.

\section{Conclusions}

The ionization balance problem has troubled astronomers in several works in the literature. Several hypotheses have been proposed to explain the differences between \ion{Fe}{ii} and \ion{Fe}{i} 
abundances but with no fully satisfactory answer yet. In this work, we propose line blending as the main reason for the overabundances of \ion{Fe}{ii}. We investigate the quality of our 
\ion{Fe}{ii} line list for unresolved blends that were missed by visual inspection from previous work. We query the VALD for lines very close to our \ion{Fe}{ii} lines and calculate how 
strong the blend is compared to our lines. We remove the ones that have significant EW contribution by setting different isolation thresholds. 

We also perform an empirical test to remove \ion{Fe}{ii} lines using high resolution spectra of 451 solar-type stars with reference stellar parameters ($T_{\mathrm{eff}}$, $[Fe/H]$, and 
microturbulence) from our previous work. Additionally, we combined two iron line lists from the literature to enhance the number of lines or improve the atomic data of the existing. 
More accurate gravities are derived based on the \textit{Gaia} parallaxes in conjunction with precise spectroscopic effective temperatures. 
Using these parameters, we exclude the \ion{Fe}{ii} lines which give abundances higher than 0.10\,dex and dispersion lower than 3\,$\sigma$. The clean \ion{Fe}{ii} line list contains 14 lines 
and is a combination of the TS13 and \cite{melendez2009} line list. The atmospheric parameters for our stellar sample with the new line list show significant improvement in the derivation of 
surface gravities, whereas the effective temperature and metallicity are in very good agreement with TS13. Moreover, the differences between \ion{Fe}{ii} -- \ion{Fe}{i} abundances do not show 
trends with the effective temperature. 

Finally, we show that the ionization equilibrium of Ti provides more accurate surface gravities than iron using for the Ti line lists of \cite{neves09} and LW13. Even though surface 
gravity still remains the parameter most difficult to constrain via spectroscopy, there are significant improvements presented in this work and in lack of parallax measurements, we can now 
provide more accurate surface gravities even for the lower main sequence stars. In this work, we propose for the optimal determination of stellar parameters, to use the TS13 line list to 
obtain $T_{\mathrm{eff}}$ and $[Fe/H]$, and the Ti line lists for a better determinations of $\log g$. This procedure will be implemented for public use in \texttt{FASMA} via the webpage.

The high quality spectra and reference atmospheric parameters of this sample can function very well to test the quality of a line list. With this data, we present an empirical solution 
to the ionization balance problem in cool stars. The bad lines excluded from this test can be attributed not only to blending effects but to other processes such as bad normalization, 
wrong atomic data, and we cannot easily distinguish which of them is responsible. We note that in this work we did not test the affects of other phenomena, such as the impact of 
3D model atmospheres, non-LTE effects, and stellar activity to explain the differences between the \ion{Fe}{ii} and \ion{Fe}{i} abundances but from our analysis we expect them to 
have secondary effects on the results. 

\section*{Acknowledgments}

The authors thank the anonymous referee for the useful comments. M. T., E. D. M., V. Zh. A., N. C. S., and S. G. S. acknowledge the support from Funda\c{c}\~{a}o para a Ci\^{e}ncia e a 
Tecnologia (FCT) through national funds and from FEDER through COMPETE2020 by the following grants UID/FIS/04434/2013 \& POCI--01--0145-FEDER--007672, PTDC/FIS-AST/7073/2014 \& 
POCI--01--0145-FEDER--016880, and PTDC/FIS-AST/1526/2014 \& POCI--01--0145-FEDER--016886. E.D.M. acknowledges the support by the fellowship SFRH/BPD/76606/2011 funded by FCT (Portugal) and by 
the Investigador FCT contract IF/00849/2015/CP1273/CT0003 and in the form of an exploratory project with the same reference. V. Zh. A., N. C. S., and S. G. S. also acknowledge 
the support from FCT through Investigador FCT contracts IF/00650/2015/CP1273/CT0001, IF/00169/2012/CP0150/CT0002, and IF/00028/2014/CP1215/CT0002 funded by 
FCT (Portugal) and POPH/FSE (EC).

This research made use of the Vienna Atomic Line Database operated at Uppsala University, the Institute of Astronomy RAS in Moscow, and the University of 
Vienna. We thank the PyAstronomy and Astropy communities.

This work has made use of data from the European Space Agency (ESA) mission {\it Gaia} (\href{https://www.cosmos.esa.int/gaia}{https://www.cosmos.esa.int/gaia}), processed by the {\it Gaia}
Data Processing and Analysis Consortium (DPAC, \href{https://www.cosmos.esa.int/web/gaia/dpac/consortium}{https://www.cosmos.esa.int/web/gaia/dpac/consortium}). Funding for the DPAC has been provided by national institutions, 
in particular the institutions participating in the {\it Gaia} Multilateral Agreement.


\bibliography{bibliography} 

\bsp	
\label{lastpage}
\end{document}